\pdfoutput=1

\documentclass[aps,longbibliography, twocolumn,amsmath,amssymb,preprintnumbers,floatfix, prb,superscriptaddress]{revtex4-2}

\usepackage[utf8]{inputenc}
\usepackage[T1]{fontenc} 
\usepackage{newtxtext}
\usepackage[upint]{newtxmath}
\usepackage{microtype}
\usepackage{textcomp}
\usepackage{eucal}
\usepackage{bm}
\usepackage{siunitx}
\usepackage{lipsum}

\usepackage{enumerate}
\usepackage{amsfonts}
\usepackage{amsmath}
\usepackage{amssymb}
\usepackage{color}
\usepackage{soul}

\usepackage{graphicx}

\usepackage[colorlinks,allcolors=blue]{hyperref}
\usepackage[capitalize]{cleveref}

\definecolor{DarkRed}{rgb}{0.65,0,0}%
\definecolor{Green}{rgb}{0,0.3,0.3}
\definecolor{Purple}{rgb}{0.3,0,0.65}
\definecolor{Red}{rgb}{1,0,0}
\definecolor{Blue}{rgb}{0,0,0.85}
\definecolor{Magenta}{rgb}{1,0,1}

\newcommand{\ve}[1]{\boldsymbol{#1}}

\newcommand{\dn}{\downarrow}
\newcommand{\up}{\uparrow}
\newcommand{\ph}{\phantom{\dag}}

\newcommand{\veck}{\ve{k}}

\newcommand{\vecm}{\ve{m}}

\newcommand{\be}{\begin{equation}}
\newcommand{\ee}{\end{equation}}

\newcommand{\prlsection}[1]{\textit{#1}.\kern0.05em---\kern0.05em\ignorespaces}

\usepackage[caption=false]{subfig}

\begin{document}
\title{Superconductor-altermagnet memory functionality without stray fields}
\author{Hans Gløckner Giil}
\affiliation{Center for Quantum Spintronics, Department of Physics, Norwegian \\ University of Science and Technology, NO-7491 Trondheim, Norway}

\author{Jacob Linder}
\affiliation{Center for Quantum Spintronics, Department of Physics, Norwegian \\ University of Science and Technology, NO-7491 Trondheim, Norway}

\begin{abstract}
A novel class of antiferromagnets, dubbed altermagnets, exhibit a non-relativistically spin-split band structure reminiscent of $d$-wave superconductors, despite the absence of net magnetization.
This unique characteristic enables utilization in cryogenic stray-field-free memory devices, offering the possibility of achieving high storage densities. 
We here determine how a proximate altermagnet influences the critical temperature $T_c$ of a conventional $s$-wave singlet superconductor.
Considering both a bilayer and trilayer, we show that such hybrid structures may serve as stray-field free memory devices where the critical temperature is controlled by rotating the Néel vector of one altermagnet, providing infinite magnetoresistance.
Furthermore, our study reveals that altermagnetism can coexist with superconductivity up to a critical strength of the altermagnetic order as well as robustness of the altermagnetic influence on the conduction electrons against non-magnetic impurities, ensuring the persistence of the proximity effect under realistic experimental conditions.
\end{abstract}
\maketitle

\section{Introduction}
The intricate interplay of superconductivity and magnetism remains a focal point in modern condensed matter physics~\cite{bergeret_rmp_05, buzdin_rmp_05, bergeret_rmp_18}. Its allure stems both from a fundamental viewpoint and cryogenic technology applications such as extremely sensitive detectors of radiation and heat as well as circuit components such as qubits and dissipationless diodes.  
Whereas superconductor-ferromagnet (SC-FM) structures have been studied extensively, the interest in antiferromagnetic materials has been comparatively limited \cite{andersen_prl_06, andersen_prb_05, bell_prb_03, hubener_jpcm_02, wu_apl_13, enoksen_prb_13} up until recently \cite{johnsen_prb_21, bobkov_prb_22, rabinovich_prr_19, falch_prb_22, jakobsen_prb_20, lado_prl_18, fyhn_prb_23, fyhn_prl_23, bobkov_prl_23, chourasiaGenerationSpintripletCooper2023, erlandsen_prb_20, thingstad_prb_21, maeland_arxiv_24}.

A particularly interesting new development is antiferromagnets that break time-reversal symmetry and feature a spin-split band structure that does not originate from relativistic effects such as spin-orbit coupling. Dubbed altermagnets in the literature, these are spin-compensated magnetic systems with a huge momentum-dependent spin splitting even in collinearly ordered antiferromagnets. \textit{Ab initio} calculations have identified several possible material candidates that can host an altermagnetic state, including metals like RuO$_2$ and Mn$_5$Si$_3$ as well as semiconductors/insulators like MnF$_2$ and La$_2$CuO$_4$~\cite{hayamiMomentumDependentSpinSplitting2019a, ahn_prb_19, lopez-moreno_pccp_16, smejkal_sa_20, reichlova_arxiv_20, smejkal_prx_22}.

Superconducting memory devices with infinite magnetoresistance have been proposed~\cite{buzdin_epl_99, tagirov_prl_99} and observed~\cite{li_prl_13} using superconducting spin-valves, a trilayer configuration comprised of a central superconductor flanked by two ferromagnets. 
By exploiting the inverse proximity effect, the critical temperature $T_c$ of the superconductor can be dynamically modulated through manipulation of the relative magnetization orientations. 
In this way, $T_c$ changes up to 1 K have been reported~\cite{singh_prx_15}.
However, the property enabling the functionality of such structure via external fields is also its drawback, depending on the precise mode of operation: the magnetization. The disadvantage is the inevitable existence of a stray field surrounding the structure, which limits how closely multiple structures of this type can be packed together without disturbing each other. Therefore, finding a way to control $T_c$ in a structure without any net magnetization could offer a major advantage to the implementation of such architecture in cryogenic devices.
Recent strides in unraveling the altermagnet/superconductivity interplay encompass a spectrum of phenomena,
including studies of Andreev reflection~\cite{sun_prb_23_altermagnets, papajAndreevReflectionAltermagnetsuperconductor2023}, Majorana zero modes \cite{ghorashiAltermagneticRoutesMajorana2023}, the Josephson effect~\cite{ouassou_prl_23, zhangFinitemomentumCooperPairing2024, beenakkerPhaseshiftedAndreevLevels2023a}, and interplay with spin-orbit interaction~\cite{wei_arxiv_23}. 

We here determine the effect of the altermagnetic spin splitting on the critical temperature of an adjacent superconductor and suggest using an AM-SC-AM trilayer as a stray field--free memory device.
Commencing our study, we investigate a simple model demonstrating coexistence of altermagnetism and superconductivity and show that the altermagnetic field is detrimental to the superconducting order parameter, akin to the Pauli limit in superconductors subjected to magnetic fields~\cite{clogston_prl_62}.
Progressing to AM-SC bilayers, we unveil a modulation in the critical temperature, caused by the altermagnetic order.
This modulation is non-monotonic as a function of the altermagnetic strength, and can both suppress or increase $T_c$ compared to the normal metal case.
We explore different geometries, showing that the relative direction of the interface and the altermagnetic order parameter yield vastly different results.
A study of AM-SC-AM trilayers is then performed, highlighting the influence of the parallel or antiparallel directions of the order parameters in the two altermagnets.
Finally, we investigate the role of impurities in the altermagnetic material. 
We find that impurities do not suppress the influence of altermagnetic spin order on the itinerant electrons, making the altermagnetic proximity effect relevant even in experiments utilizing materials that have a short mean free path.

\section{Theory}
The lattice Bogoliubov--de\,Gennes (BdG) framework~\cite{zhu2016, degennes1966} is suitable for studying AM-SC heterostructures.
We employ an attractive Hubbard Hamiltonian to model a conventional phonon-mediated $s$-wave singlet superconductor:
\begin{align}
    H_\text{U} =  - \sum_{i} U_{i} c_{i\downarrow}^\dag  c_{i\uparrow}^\dag c_{i \uparrow} c_{i \downarrow},
\end{align}
where $c_{i\sigma}$ and $c^\dag_{i\sigma}$  destroy and create an electron with spin $\sigma$, and $U_i>0$ is the magnitude of the attractive potential. The attractive Hubbard $U$-term generally differs from the effective electron-electron interaction mediated by phonons (obtained by performing a Schrieffer-Wolff transformation) in terms of momentum and spin indices. Nevertheless, at the mean-field level these models give the same result for the conventional BCS channel for Cooper pairing.
We define $b_i = c_{i \downarrow} c_{i \uparrow}$ and perform a mean-field expansion $b_{i} = \delta b_i + \langle b_i \rangle$, ignoring second-order terms in the deviation from expectation values. Finally, we define a (site-dependent) superconducting order parameter
\begin{align}
\label{eq:gapeeq}
    \Delta_i =   U_{i} \langle c_{i \uparrow} c_{i \downarrow} \rangle,
\end{align}
and arrive at the mean-field Hamiltonian
\begin{align}
    H_\text{mf} = -\sum_{i} (\Delta_i^{\vphantom{*}} c^\dag_{i \downarrow} c^\dag_{i\uparrow} 
    + \Delta_i^* c_{i\uparrow} c_{i\downarrow}), 
\end{align}
where we disregarded a constant term which is absorbed in the ground state energy.

We note that a constant on-site potential $U_i = U$ corresponds to an isotropic gap in momentum-space that pairs electrons with opposite spin and momentum, which is consistent with the Bardeen-Cooper-Schrieffer theory for an $s$-wave superconductor.
We employ this mean-field superconducting Hamiltonian, including also the effect of altermagnetism and impurities,
\begin{equation}
\label{eq:Hn}
    \begin{split}
        H =&  E_0
		- \sum_{i\sigma} ( \mu - w_i) c^\dag_{i\sigma} c_{i\sigma}
  - \sum_i (\Delta_i^{\vphantom{*}} c^\dag_{i\downarrow} c^\dag_{i\uparrow} 
    + \Delta_i^* c_{i\uparrow} c_{i\downarrow}) 
    \\& 
		- \sum_{\langle i, j \rangle \sigma} t_{ij} c^\dag_{i\sigma} c_{j\sigma}
		- \sum_{\langle i, j \rangle \sigma\sigma'} (\bm{m}_{ij} \cdot \bm{\sigma})_{\sigma\sigma'} c^\dag_{i\sigma} c_{j\sigma'} ,
    \end{split}
\end{equation}
where $\mu$ is the chemical potential,  $\bm{\sigma} = (\sigma_1, \sigma_2, \sigma_3)$ is the Pauli vector, and $w_i$ is an impurity potential taken to be randomly distributed at a given number of sites in the magnet.
For comparison, we consider two different forms of the spin-dependent interaction $\vecm_{ij}$: i) an on-site potential $\vecm_{ij} = m_z \delta_{ij} \hat z $, corresponding to a ferromagnetic term, and ii) $\bm{m}_{ij} = +m\bm{e}_z$ for nearest-neighbor hopping along the $x$ axis and $\bm{m}_{ij} = -m\bm{e}_z$ for hopping along the $y$ axis, corresponding to an effective altermagnetic term, similar to what was used in Ref.~\onlinecite{ouassou_prl_23}.
The spin-dependent hopping term parametrized by $\vecm_{ij}$ in our Hamilton-operator can be understood as a Coulomb-exchange interaction experienced by electrons that are hopping on top of a background of localized spins that form collinear antiferromagnetic order \cite{reichlova_arxiv_20}.
For a bulk altermagnet, the spin-dependent hopping used this article as well as in Ref.~\onlinecite{ouassou_prl_23} takes exactly the cosine form suggested discussed in Refs.~\onlinecite{reichlova_arxiv_20, am_emerging, smejkalGiantTunnelingMagnetoresistance2022}, when Fourier transforming to momentum space.
Moreover, both the Hubbard model \cite{dasRealizingAltermagnetismFermiHubbard2023, nakaSpinCurrentGeneration2019} and the tight-binding model with spin coupling between itinerant and localized spins have recently used \cite{brekkeTwodimensionalAltermagnetsSuperconductivity2023} to derive dispersions that qualitatively match the low-energy form of the cosine Hamiltonian.
The effect of the antiferromagnetic order in the localized spins exerted on the itinerant fermions $\{c_{i\sigma}, c_{i\sigma}^\dag\}$ in the Hamiltonian Eq.~\eqref{eq:Hn} is thus modelled via $\vecm_{ij}$, owing to the fact that this is an effective model for the conduction electrons which breaks the $PT$-symmetry required to have a spin-split altermagnetic band structure. 
Formally, one could in principle solve self-consistently for the parameter $m$ in the altermagnet to determine how it is affected by superconductivity. This would be relevant for spontaneous and intrinsic coexistence of altermagnetism and superconductivity in the same material. Instead, we take $m$ to be a fixed constant and solve for $\Delta$ self-consistently.
This scenario is experimentally relevant in a scenario where the altermagnetic spin-splitting has been induced by placing a thin superconductor on top of an altermagnet, stacking them along the $z$-direction, whereas the spin-splitting of the bands is present in the $k_x-k_y$ plane. This is similar to what has been done experimentally in thin ferromagnetic insulator/superconductor systems \cite{li_prl_13, kolenda_observation_2016, rouco_prb_19}, and obviates the need to solve self-consistently for the spin-splitting parameter as it is induced from an external source. Note that in such a case, the induced $m$ in the superconductor can be smaller than $\Delta$ despite the altermagnetic spin-splitting in the host material being much larger than $\Delta$. This is because the induced spin-splitting via the proximity effect scales with the tunnel coupling to the material, which strongly suppresses its magnitude.

We assume nearest neighbor hopping, i.e. $t_{ij} = t$, and scale all other parameters in units of $t$.
The superconducting order parameter is determined from the site-dependent self-consistent gap equation in Eq.~\eqref{eq:gapeeq}.
We assume here that the attraction only occurs in the singlet channel. The singlet phase is more robust than the triplet channel when impurity scattering, which is always present to some extent in real materials, is included. We note that even if the pairing potential only exists in the singlet channel, triplet superconducting correlations can still be induced via the proximity effect when such a superconductor is in contact with an altermagnet. For a discussion concerning the pairing potential in the triplet channel when superconductivity coexists with altermagnetism, see Ref. \onlinecite{zhuTopologicalSuperconductivityTwoDimensional2023a}.
Throughout the paper, we fix $\mu = -t/2$ and $U_i = 1.7t$.
The magnetic terms, superconducting order parameters, and impurity potentials are only nonzero in their respective regions. Specifically, the altermagnetic term $\vecm_{ij}$ is finite only when both sites $i, j$ inside are in the altermagnet. Expectation values
of physical observables are formally computed by performing a trace (using a complete basis set) over the density matrix $\rho$ and the matrix-representation of the observable under consideration. The details of this
density matrix are not needed for the results presented in our work. This is because the superconducting order parameter and other expectation values
can in practice be obtained self-consistently without explicitly computing $\rho$ first, the reason being that expectation values of creation-annihilation pairs of the diagonalized quasiparticle operators give Fermi-Dirac distribution functions.

\section{Methodology}
At each site~$i$, the fermionic operators can be organized into Nambu vectors 
$\hat{c}_i \equiv (c_{i \up}^{\ph}, 
c_{i\dn}^{\ph}, c_{i\up}^\dag, c_{i\dn}^\dag)$
, which may in turn be collected into a $4N$-element vector $\check{c} \equiv (\hat{c}_1, \ldots, \hat{c}_N)$ encompassing all fermionic lattice operators.
The Hamiltonian operator is subsequently represented using a $4N\times4N$ matrix: $H = E_0 + \frac{1}{2} \check{c}^\dag \check{H} \check{c}$.
We solve the BdG equation by numerically diagonalizing $\check H$ and expressing physical observables such as the superconducting gap in Eq.~\eqref{eq:gapeeq} in terms of its eigenvectors and eigenvalues. 
This process entails an initial guess $\Delta_g$ for the order parameter, and then self-consistently diagonalizing the Hamiltonian until the superconducting gap equation converges. In this paper, however, our main interest is the critical temperature $T_c$, and thus we do not need the explicit numerical value of the gap. With that said, we typically find a low-temperature gap magnitude of order $\Delta \simeq 0.15t$ with our parameters.
Instead, we perform $N_\Delta$ self-consistent iterations and compare the resulting value of the order parameter with the small initial value $\Delta_g = 10^{-4} t$.
This solution strategy is very similar to the methodology used, e.g., in Ref.~\onlinecite{johnsen_prl_20}.
We define the SC as being in the superconducting state when the median value of the order parameter inside the superconductor has increased compared to the initial value $\Delta_g$.
The critical temperature is subsequently ascertained by performing a binomial search in critical temperatures, as was done in Ref.~\onlinecite{ouassou2019}. In order to make the computational time manageable, it is necessary to consider a lattice size that is much smaller than in an experimental setting. For instance, in order to ensure that the width of the superconducting layer is comparable to the superconducting coherence length, which is inversely proportional to $\Delta$, one must use a large value for the superconducting order parameter. Nevertheless, the BdG lattice framework is known to give predictions that compare well, both qualitatively and quantitatively, with experimentally realistic systems \cite{blackschaffer_prb_10, english_prb_16} as long as the ratio of the length scales in the problem (such as the width of the system and the coherence length) is reasonable, which is the approach we have taken.

\section{Results and discussion}

\subsection{Altermagnetic destruction of the superconducting order}

\begin{figure}[t!]
    \centering
    \subfloat[]{\includegraphics[width = 0.65 \linewidth]{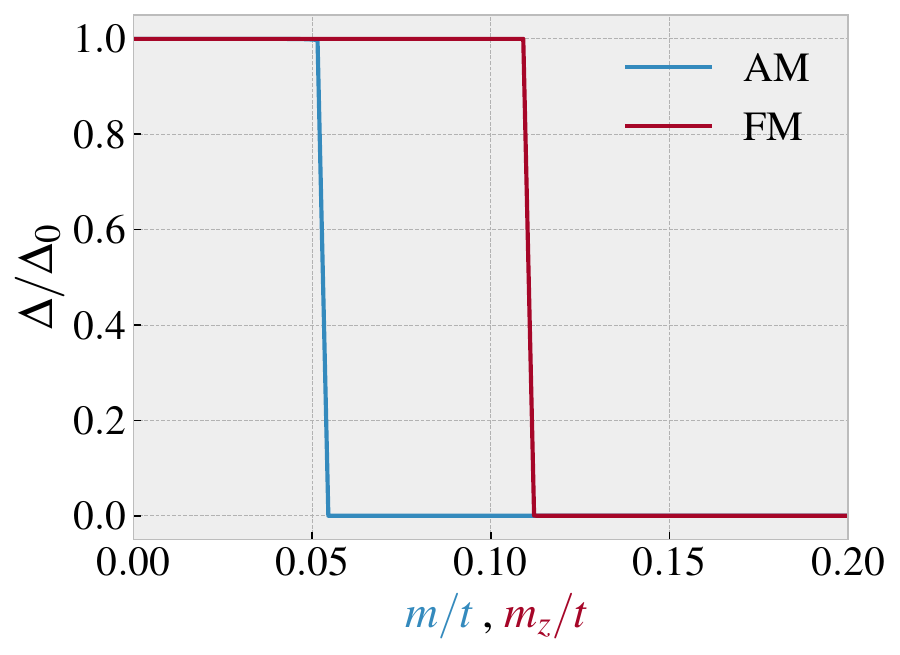}}\\
    \subfloat[]{
    \includegraphics[width = 0.65 \linewidth]{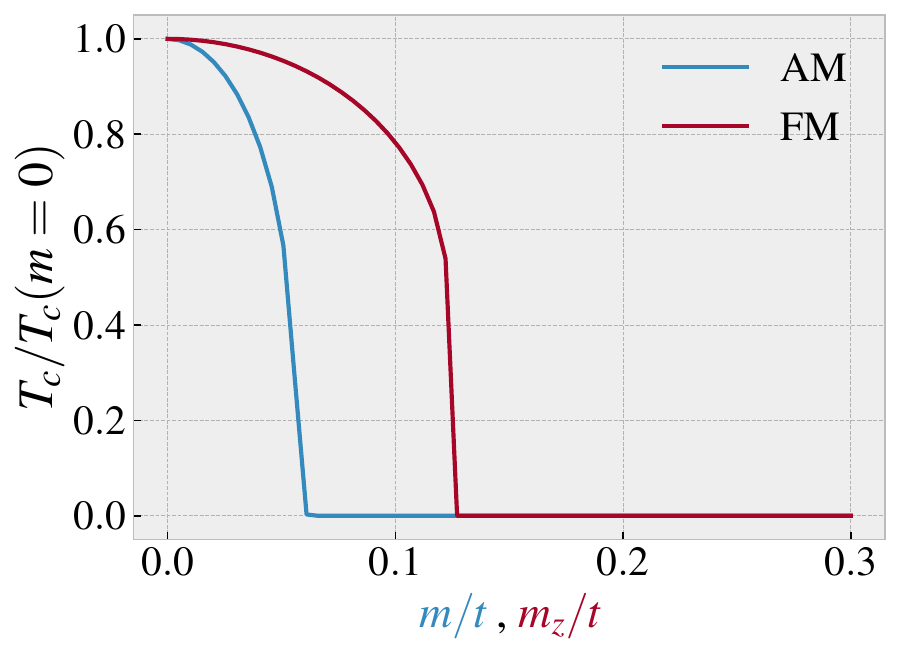}
    }
    \caption{
    The (a) order parameter and (b) critical temperature as a function of the alter- and ferromagnetic strength in a $N_x = N_y = 20 a_0$ structure with coexisting superconductivity and altermagnetic/ferromagnetic spin splitting. The termperature in (a) is set to $T = 0.01 t / k_B$, where $k_B$ is the Boltzmann constant.
    }
    \label{fig:coex}
\end{figure}

Prior to delving into heterostructures of superconductors and altermagnets, it is instructive first to understand the effects of the altermagnetic term in Eq.~\eqref{eq:Hn} on the superconducting order, and employ periodic boundary conditions along both axes.
To this end, we consider a system with coexisting altermagnetic and superconducting order. We vary the altermagnetic strength $m$ and calculate the critical temperature self-consistently using the methodology outlined above.
The results are shown in Fig.~\ref{fig:coex}: the effect of altermagnetism is to suppress the superconducting order, which vanishes for an altermagnetic strength of $m \approx 0.05 t$.
The results are juxtaposed with the effects of ferromagnetism, which also suppresses the superconductivity in a similar way, as is well-known \cite{clogston_prl_62, chandrasekhar_apl_62}, although the critical field is much larger than in the altermagnetic case.
We note that in general there exists additional solutions to the self-consistency equation besides the one shown in the upper pane of Fig.\ \ref{fig:coex}, such as $\Delta = 0$. These solutions have a higher free energy than the solution for $\Delta$ that we have presented. Thus, we are presenting the solution for $\Delta$ which corresponds to the thermodynamic ground-state of the system. These unstable branches are discussed in detail in the Appendix of Ref. \onlinecite{ouassouVoltageinducedThinfilmSuperconductivity2018a}. 

After establishing the analogous interaction between the altermagnetic and ferromagnetic terms with superconductivity, a natural inquiry arises regarding the impact of altermagnets on superconductors within heterostructures. Specifically, we shall focus our attention on the influence of altermagnets on the critical temperature within AM-SC systems.

\subsection{Junction geometries}
\begin{figure}[t!]
    \centering
    \captionsetup[subfloat]{position=top,labelformat=empty,  font ={scriptsize, bf}}
    \subfloat[]{
    \includegraphics[width=0.6\linewidth]{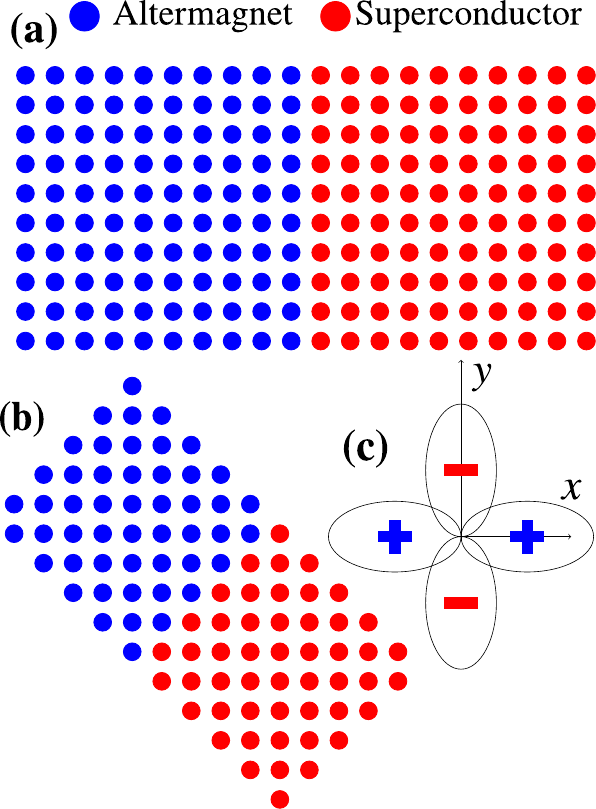}}
    \\
    \subfloat[(d)]{\includegraphics[width = 0.65 \linewidth]{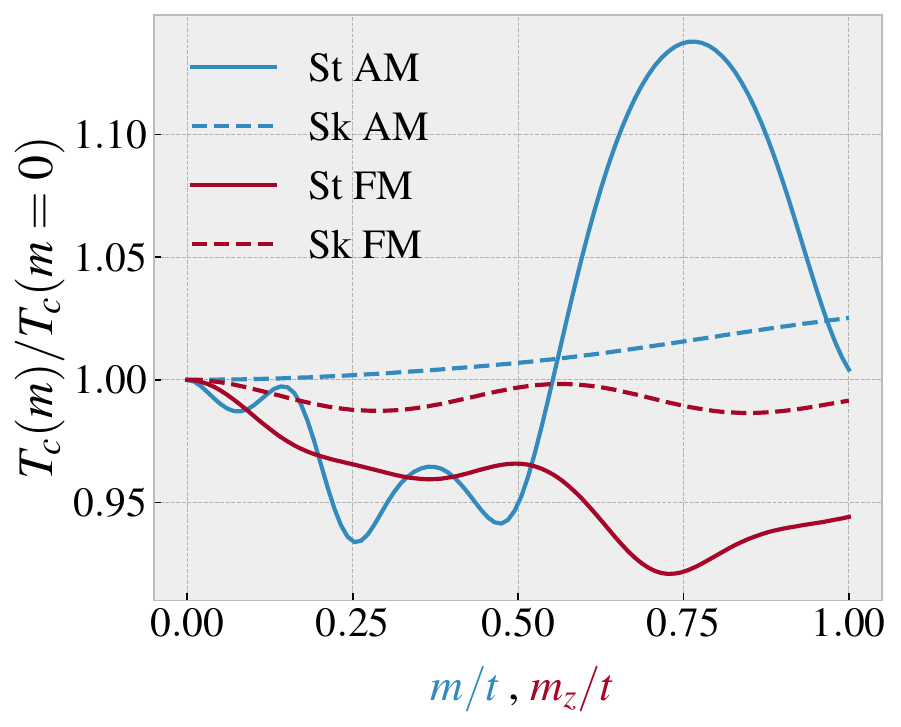}
    \label{fig:geometries_d}
    }
    \caption{(a) The straight junction and (b) the skewed junction geometries. (c) the hopping term for a spin-up (spin-down) electron (hole). For a spin-down (spin-up) electron (hole), the signs are reversed.
    In (d): the critical temperature for the straight (St) and skewed (Sk) geometries with $N_y = 20 a_0$, $ N_x^{AM} = 10 a_0$,
     $N_x^{SC}=6 a_0$, and $N_\Delta = 50$.
    }
    \label{fig:geometries}
\end{figure}

Employing a square lattice with lattice constant $a_0$, we explore two distinct AM-SC geometries: a straight junction, where the interface is aligned with the crystallographic axis, and a skewed junction, where the interface is rotated $45^\circ$ compared with the crystallographic axis, see Fig.~\ref{fig:geometries}.
Here, and in the rest of the paper, we employ periodic boundary conditions along the axis parallel to the interface, and hard wall boundary conditions along the axis perpendicular to the interface.
We denote the number of lattice sites in the $x$ ($y$) direction by $N_{x(y)}$.
In the straight junction, hopping across the interface happens exclusively along the $x$-axis, while in the skewed junction, hopping across the interface happens equally along the $x$- and $y$- axis. 
The inverse proximity effect in the SC-AM system can be probed by calculating the critical temperature $T_c$ in the SC. The result is depicted in Fig.~\ref{fig:geometries}, highlighting a significant disparity in the behavior between the straight and skewed junction configurations. 
We emphasize that the normalization in this figure is the critical temperature in a superconductor in contact with a $m=0$ altermagnet, i.e. a normal metal. Hence, the critical temperature is never raised compared to the critical temperature in the bulk superconductor, but it is raised compared to the critical temperature in a superconductor in proximity with a normal metal. 
In the skewed junction, the effect of the altermagnetism is to suppress Andreev reflection \cite{andreev_jetp_64}, which is the underlying mechanism causing the (inverse) proximity effect. This happens because the altermagnetic term causes different hopping for electrons and holes involved in Andreev reflection.
Thus, the inverse proximity effect is suppressed for high values of $m$, causing the critical temperature to increase.
In the case of the straight junction, an additional factor comes into play—induced magnetization brought about by the inverse proximity effect \cite{linder_np_15}. This phenomenon leads to a pronounced oscillatory behavior in the critical temperature.
The induced magnetization can be understood by noting that spin-up electrons favor hopping in the $x$-direction, causing leaking spin-up electrons from the SC into the AM to be trapped in the AM for large $m$. 
In the skewed junction, this effect is averaged out, and the induced magnetization in the SC vanishes.
\begin{figure}[t!]
    \centering
    \captionsetup[subfloat]{position = bottom,labelfont=bf}
    \subfloat[]{\includegraphics[width = 0.6 \linewidth]{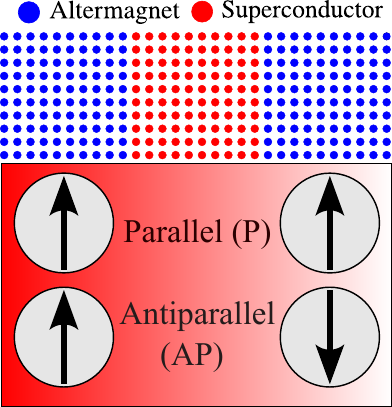}}
    \\
    \subfloat[]{
    \includegraphics[width = 0.65 \linewidth]{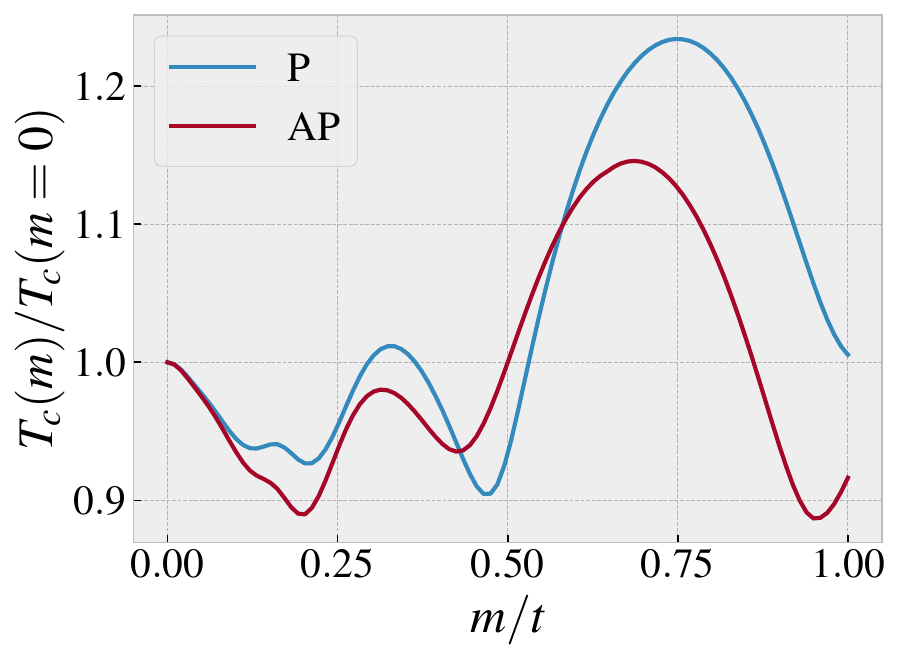}
    \label{fig:PAP_c}
    }
    \caption{In (a): the AM-SC-AM system with P and AP alignment of the altermagnets. In (b): the critical temperatures in a system where the length of the SC is $12 a_0$, and $N_y = 20 a_0$. The altermagnets on either side have a length of $10 a_0$. 
    }
    \label{fig:PAP}
\end{figure}
\begin{figure}[htb]
    \centering
    \captionsetup[subfloat]{position = bottom,labelfont=bf}
    \subfloat[]{
    \includegraphics[width = 0.48\linewidth]{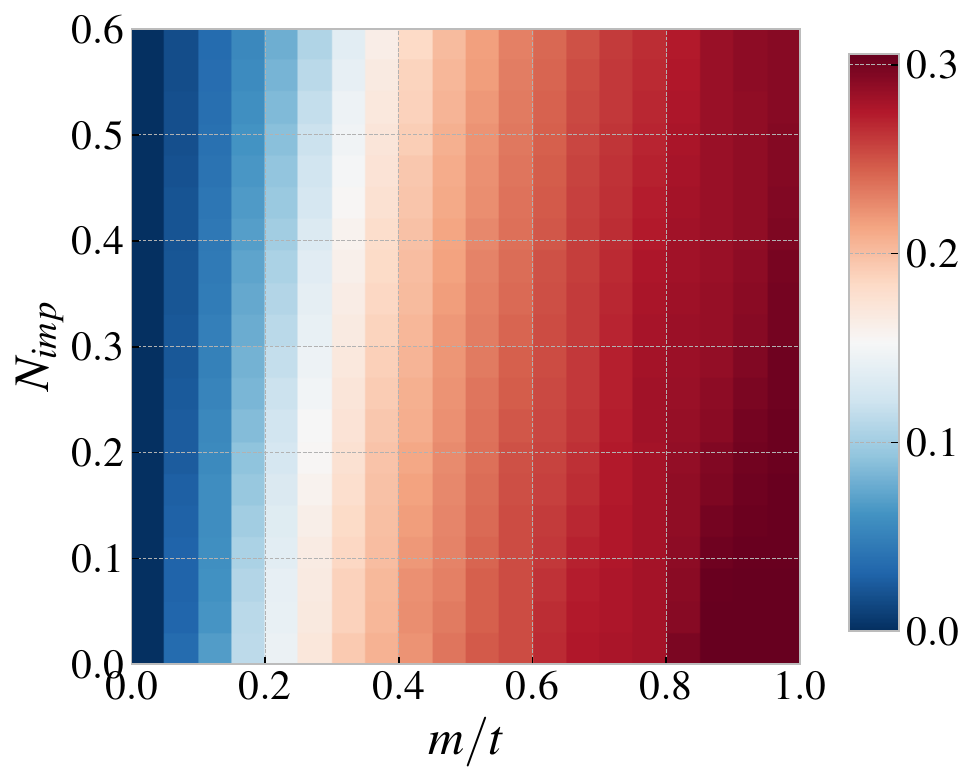}
    \label{fig:imp_am_a}
    }
    \subfloat[]{\includegraphics[width = 0.48 \linewidth]{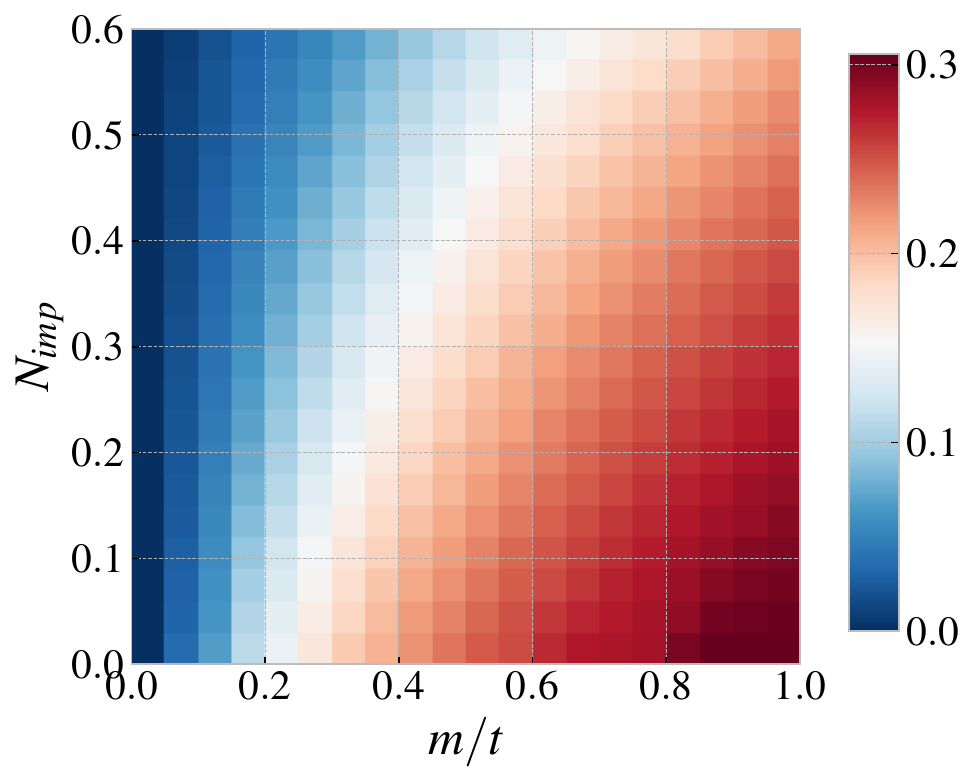}
    \label{fig:imp_am_b}
    }
    \\
    \subfloat[]{\includegraphics[width = 0.48 \linewidth] {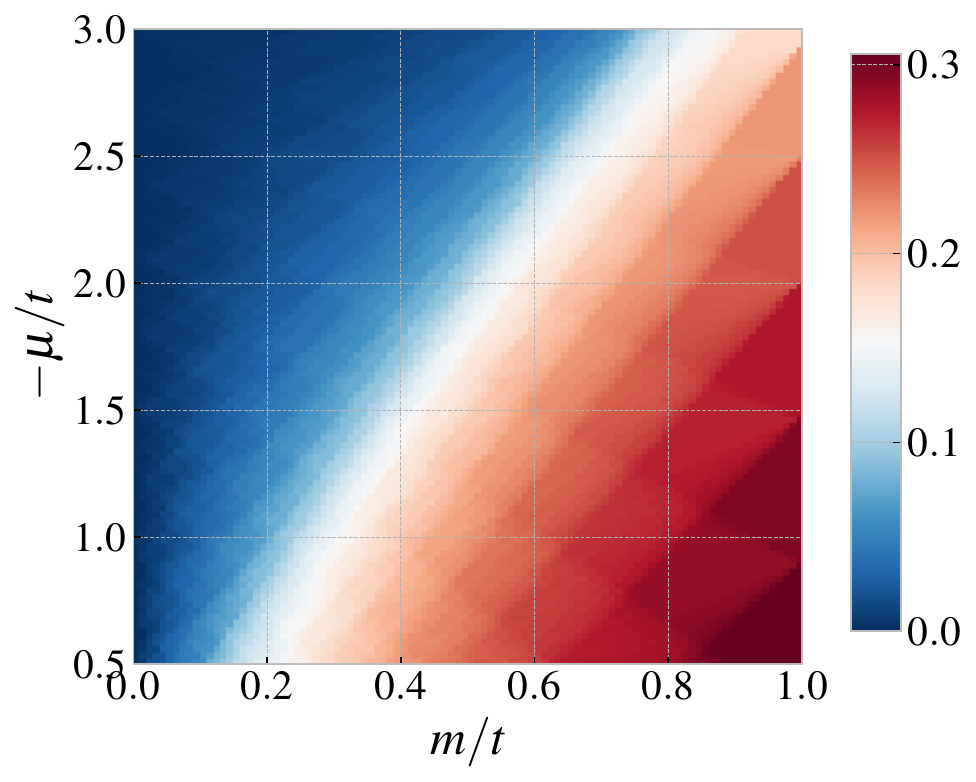}
    \label{fig:imp_am_c}
    }
    \caption{$\Delta C_\uparrow$ as a function of $m$ and $N_i$, for impurity strength of (a) $w_i = 1 t$ and (b) $w = 3 t$. 
    (c) $\Delta C_\uparrow$ as a function of $m$ and $\mu$ in a system without impurities.,
    The system size is $N_x = N_y = 20 a_0$.}
    \label{fig:imp_am}
\end{figure}

\subsection{AM-SC-AM trilayers}
An intriguing extension to the discussion above can be achieved by adding another altermagnet to the AM-SC system considered above.
This system entails two distinct scenarios: one where the two altermagnets are aligned and one where the second altermagnet is rotated (in real space) by $90 ^ \circ$.
We refer to these situations as a parallel (P) and antiparallel (AP) alignment, see Fig.~\ref{fig:PAP}.
Rotating the second altermagnet is akin to changing the sign of $m$ in this region, or equivalently to a $180^\circ$ rotation in spin space.
In Fig.~\ref{fig:PAP_c}, the critical temperature of the SC is calculated for different values of $m$ in the two different systems.
In the P alignment, the situation is analogous to the AM-SC system considered above, and we see a similar $T_c$ modulation pattern.
In the AP alignment case, the critical temperature is lower than in the P alignment for most values of $m$.
To understand why this is the case, we note that the superconducting coherence length in our system $\xi_S = \hbar v_F/ \pi \Delta_0$, where $v_F = \langle |dE_{{k}}/d{\veck}| \rangle / \hbar$ is the normal state Fermi velocity, where $\langle \ldots \rangle$ represents averaging over the Fermi surface, which can be calculated by introducing periodic boundary conditions along both axes \cite{johnsenControllingSuperconductingTransition2020a}, is comparable to the system length. This is typically the regime investigated experimentally.
In light of this, we attribute the lower critical temperature to the appearance of crossed Andreev reflection (CAR), sometimes referred to as nonlocal Andreev reflection~\cite{falci_europhys_01}.
It is well known that for an F-S-F heterostructure, the AP alignment of ferromagnets gives enhanced CAR compared to the P alignment~\cite{deutscher_apl_00}, and we attribute the results of Fig.~\ref{fig:PAP_c} to a similar origin. The CAR process (strictly speaking, inverse CAR) breaks up a Cooper pair into electrons that become spatially separated in different leads, thus suppressing the superconducting condensate. As more Cooper pairs are transmitted out of the SC due to CAR, the critical temperature drops accordingly. 
Importantly, switching between the P and AP alignment can experimentally be performed by rotating the Néel vector, since this effectively switches the spin-up and spin-down bands in the altermagnet. A similar {$T_c$ modulation} in conventional antiferromagnets was very recently reported \cite{kamraCompleteSuppressionNeel2023}.
Notably, the Néel vector has been found to be controllable by spin transfer torques~\cite{cheng_prb_15, urazhdin_prl_07}, spin-orbit torques \cite{xu_jap_23}, and by optical methods~\cite{grigorev_acs_22}.
This opens the possibility of using the suggested device as a stray field--free memory device operating in the THz regime, enabling the prospect of ultrafast switching.

\begin{figure}[t!]
    \centering
    \includegraphics[width = 0.65 \linewidth]{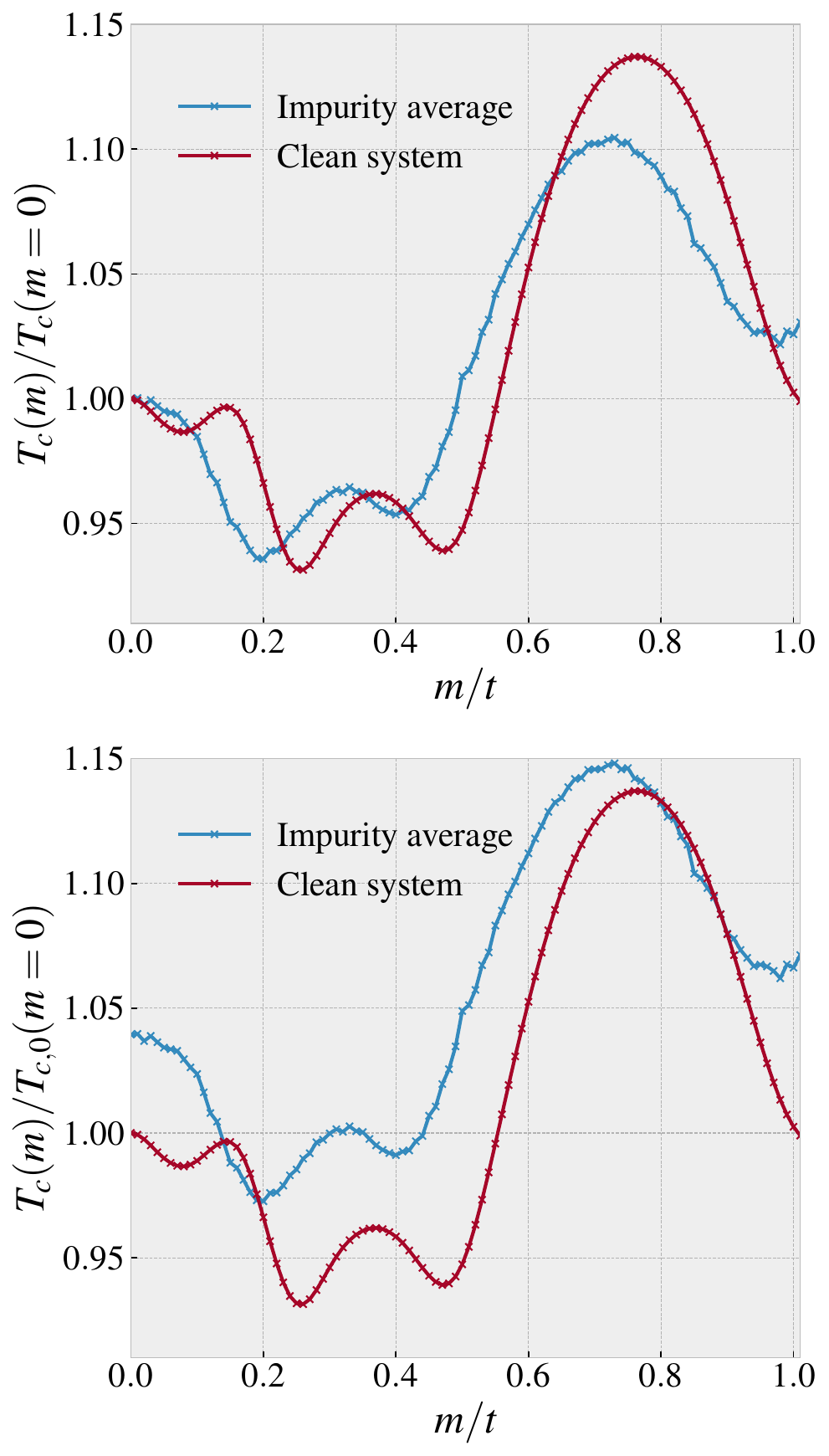}
    \caption{The impurity average plotted together with the clean system, which is the same as the straight system in Fig.~\ref{fig:geometries_d}. 
    In the lower pane, the temperatures are normalized to the zero-impurity and zero-magnetism critical temperature $T_{c,0}(m=0)$, showing that $T_c$ is slightly higher in the presence of impurities.
    In the upper pane, the temperatures are normalized to unity for $m=0$, i.e. the two curves are normalized by a different factor, which illustrates that the variation in $T_c$ with $m$ is of similar magnitude in both cases.}
    \label{fig:imps}
\end{figure}

\subsection{Impurity scattering}
Materials with substantial impurity scattering are highly relevant for experiments. 
For this reason, we will concentrate on the role of impurities in altermagnets, before moving on to the proximity effect in a system with a dirty altermagnet.
Impurities are accounted for through an on-site potential~\cite{asano_prb_07, asano_prl_07, li_npj_21} at a fraction $N_i$ of all sites, with a fixed strength $w_i$, and randomly chosen sites in the altermagnet, similar to the methodology in Ref.~\onlinecite{li_npj_21}.  Observables are calculated by averaging over $100$ different impurity configurations. 
As the impurity scattering is isotropic, one might expect that the altermagnetic spin-splitting, which is anisotropic, disappears in the presence of impurities.
To test this, we define the \textit{bona fide} order parameter $\Delta C_\uparrow$,
\begin{align}
    \Delta C_\uparrow = \sum_i \left[ \langle c_{i \uparrow}^\dagger c_{i+ \hat x \uparrow} \rangle 
    -  \langle c_{i \uparrow}^\dagger c_{i+ \hat y \uparrow} \rangle \right],
\end{align}
which is a measure of the anisotropy of the effective hopping parameter (for spin-up particles) in the system, and depends on both $t$ and $m$ in general.
For square systems, we expect the system to be invariant under $C_4$ rotations for $m=0$, i.e. $\Delta C_\uparrow =  0$.
Thus, we can use $\Delta C_\uparrow$ to determine whether the system is altermagnetic or not.
In Fig.\ \ref{fig:imp_am}, we plot the results for different values of $m$ and $N_i$, for $w_i = 1.0t$ and $w_i = 3.0t$, comparable to other values used in the literature~\cite{asano_prl_07}. 
Evidently, the altermagnetic order is resilient to the non-magnetic impurities in the system; the slight suppression of the order parameter for strong impurity scattering (i.e. the upper parts of the plots in Fig.~\ref{fig:imp_am_b} and Fig \ref{fig:imp_am_a}) can be explained by the fact that the impurities takes the form of local chemical potentials. Hence, a high concentration of impurities has the effect of an effective renormalized global chemical potential. To show this, we calculate the order parameter $\Delta C_\uparrow$ for different values of  global chemical potentials, as shown in Fig.~\ref{fig:imp_am_c}. The suppression of the order parameter for high absolute values of $\mu$ is similar to the behavior of the order parameter for high impurity configurations in Fig.\ \ref{fig:imp_am_a} and \ref{fig:imp_am_b}. The reason for the dependence of the order parameter magnitude on $\mu$ is that the latter determines the filling factor and density of states at the Fermi level, which in turn affects the superconducting pairing.

Finally, we repeat the calculations of the straight junction in Fig.~\ref{fig:geometries}, including impurities in the AM.
We set the strength of the impurities to $w_i = 1.0t$ and the fraction of sites occupied by impurities to $0.2$, and perform the calculations for $100$ different impurity configurations, before averaging over the resulting values of the critical temperature. 
The results are shown in Fig.~\ref{fig:imps}, and show that although the critical temperature curve is different from the clean system, the (inverse) proximity effect is still present, which is evident from the fact that the critical temperature varies with a similar magnitude compared with the clean system. 

Thus, we conclude that impurities are not strongly detrimental to the altermagnetic modulation of the superconducting order, which means that the effect should be experimentally visible even for dirty materials.
We have thus found that the analogy between $d$-wave superconductor and altermagnetism~\cite{smejkal_prx_22} is not useful in impurity considerations: whereas $d$-wave superconductivity is highly sensitive to non-magnetic impurities, the altermagnetic effect on the conduction electrons survives even in the presence strong impurity potentials. This is due to the ''$d$-wave'' Fermi surface of altermagnets being spin-split, prohibiting the scattering between the spin bands in the absence of spin-flip impurities.

\section{Conclusion}
We have solved the lattice Bogoliubov-de Gennes equations in heterostructures of superconductors and altermagnets.
Our study indicates that altermagnetic materials have the potential to be used in cryogenic spintronic devices, for instance as stray-field-free spin switches showing infinite magnetoresistivity, using spin-transfer torques, spin-orbit torques, or optical methods to rotate the Néel vector.
Non-magnetic impurities are not severely detrimental to the altermagnetic proximity effect, allowing for this effect to be present also in altermagnetic materials in the diffusive transport regime.

\begin{acknowledgments}
We thank J. A. Ouassou, E. W. Hodt, and B. Brekke for helpful comments and discussions.
This work was supported by the Research
Council of Norway through Grant No. 323766 and its Centres
of Excellence funding scheme Grant No. 262633 “QuSpin.” Support from
Sigma2 - the National Infrastructure for High Performance
Computing and Data Storage in Norway, project NN9577K, is acknowledged.
\end{acknowledgments}

\bibliography{main}

\end{document}